\documentclass[aps, prl, amsfonts, amssymb, amsmath, floatfix, superscriptaddress, showpacs, reprint]{revtex4-1}
\usepackage{times}
\usepackage{bm}
\usepackage{hyperref}
\usepackage{graphicx}
\usepackage{epsfig} 
\usepackage{epstopdf}
\usepackage{color}
\usepackage{soul}

\begin{document}

\title{The Structure of Plasma Heating in Gyrokinetic Alfv\'enic Turbulence}
\author{Alejandro \surname{Ba\~n\'on Navarro}}
\email{banon@physics.ucla.edu}
\affiliation{Department of Physics and Astronomy, UCLA, 475 Portola Plaza, Los Angeles, CA 90095-1547, USA}
\author{Bogdan Teaca}
\email{bogdan.teaca@coventry.ac.uk}
\affiliation{Applied Mathematics Research Centre, Coventry University, Coventry CV1 5FB, United Kingdom}
\author{Daniel Told}
\affiliation{Department of Physics and Astronomy, UCLA, 475 Portola Plaza, Los Angeles, CA 90095-1547, USA}
\author{Daniel Groselj}
\affiliation{Department of Physics and Astronomy, UCLA, 475 Portola Plaza, Los Angeles, CA 90095-1547, USA}
\affiliation{Max-Planck-Institut f\" ur Plasmaphysik, Boltzmannstra\ss e 2, D-85748 Garching, Germany}
\author{Paul Crandall}
\affiliation{Department of Physics and Astronomy, UCLA, 475 Portola Plaza, Los Angeles, CA 90095-1547, USA}
\author{Frank Jenko}
\affiliation{Department of Physics and Astronomy, UCLA, 475 Portola Plaza, Los Angeles, CA 90095-1547, USA}
\begin{abstract}
We analyze plasma heating in weakly collisional kinetic Alfv\'en wave turbulence using high resolution gyrokinetic simulations spanning the range of scales between the ion and the electron gyroradii. Real space structures that have a higher than average heating rate are shown not to be confined to current sheets. This novel result is at odds with previous studies, which use the electromagnetic work in the local electron fluid frame, i.e.~$\mathbf{J} \!\cdot\! (\mathbf{E} + \mathbf{v}_e\times\mathbf{B})$, as a proxy for turbulent dissipation to argue that heating follows the intermittent spatial structure of the electric current. Furthermore, we show that electrons are dominated by parallel heating while the ions prefer the perpendicular heating route. We comment on the implications of the results presented here. 
\end{abstract}
\maketitle

{\em Introduction.---} The radial temperature profile of the solar wind, as measured by spacecraft, can only be explained by the presence of heating throughout the heliosphere~[\onlinecite{richardsongrl03}]. Identifying the physical mechanisms that dissipate small-scale turbulent fluctuations, ultimately converting the turbulent energy into heat, is thus one of the major unsolved problems in the solar wind community~[\onlinecite{brunolrsp13}]. Due to the low collisionality of the solar wind, a kinetic model is necessary to understand the effects that contribute to the heating of the plasma. Over the past years, a major effort has been put into computational studies of the problem using fully kinetic and reduced kinetic models in two or three spatial dimensions~\cite{Verscharen2012,  Chang2014, Servidio2015,  toldprl15, parasharpop11, TenBargeapj13, Karimabadipop13, wanprl12, vasquez_taj14, hughes_grl14, wanprl15, Leonardis2016, gary_taj16}.
 
Recent kinetic plasma simulations found that the dissipation of electromagnetic turbulent fluctuations occurs in an intermittent hierarchy of coherent structures~\cite{parasharpop11, TenBargeapj13, Karimabadipop13, wanprl12, wanprl15, Leonardis2016}, confirming a series of solar wind satellite observations~\cite{retinonp07, sundkvistprl07, osmanajl11, osmanprl12, perriprl12,perrone_taj16}. This led to the conjecture that heating in turbulent space plasmas is also highly inhomogeneous, patchy and occurs predominantly in current sheets. However, to date, fully kinetic simulations in two or three spatial dimensions do not compute the actual local heating rate, depending instead on a number of proxies~\cite{wanprl12, wanprl15}. The electromagnetic work on the plasma particles in the fluid frame represents the most common proxy for heating in a weakly collisional turbulent plasma.

This work addresses for the first time in a direct way the real space structure of heating in weakly collisional kinetic Alfv\'en wave (KAW) turbulence, which has been demonstrated to be a crucial ingredient of solar wind turbulence~\cite{podestasp13, Boldyrev2015, Vasconez2015, howes_pta15}.  In this context, a series of questions emerge. Does the plasma heating occur predominantly in current sheets or not? How much of the structure exhibited by  collisional heating can be captured by electromagnetic work and how appropriate is the latter as a dissipation proxy for turbulent heating? Is the heating more homogeneous, or does it have a patchy nature? What are the main mechanisms for turbulent heating and do they occur predominantly in the perpendicular or parallel direction in velocity space? We provide answers to these questions before commenting on the implications of the results found in the broader context of solar wind research.

{\em Simulation setup.---} In this work, KAW turbulence is studied using the gyrokinetic (GK) theory~\cite{brizardrmp07}, which is a rigorous limit of kinetic theory in strongly magnetized plasmas. It assumes low frequencies (compared to the ion cyclotron frequency) and small fluctuation levels. Therefore, it excludes cyclotron resonances and stochastic heating.  A recent comparison [\onlinecite{toldnjp16}] with the fully kinetic model found that GK is able to accurately reproduce the physics of KAWs for the parameters considered in this study. Furthermore, the reduction of the problem to a five-dimensional phase space allows for the use of a realistic mass ratio and a grid-based numerical scheme, which does not suffer from discrete particle noise and is able to treat heating explicitly via a well-defined collision operator. To date, several previous works have considered the plasma heating problem in three-dimensional  turbulence using GK theory~\cite{TenBargeapj13, toldprl15}. However, none of these studies has measured directly the collisional dissipation in real space. 

\begin{figure}[t]
\centering
\includegraphics{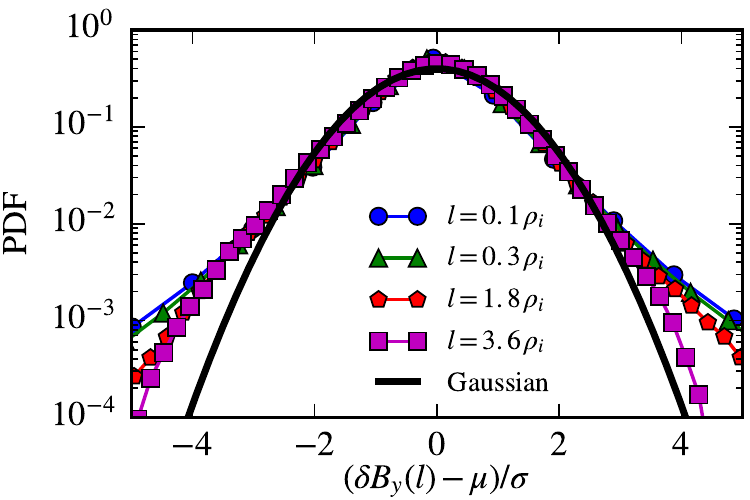}
\caption{(Color online) PDFs of magnetic field increments for different separation lengths $l$. The increments are normalized by 
their standard deviations ($\sigma$) and the averages ($\mu$) were subtracted before computing the PDFs.}
\label{structure}
\end{figure}

The data used in this Letter is taken from the simulation presented in Ref.~\cite{toldprl15}, and it is briefly summarized in the following. The nonlinear gyrokinetic system of equations is solved with the Eulerian code {\sc Gene}~\cite{genecode}, capturing the dynamics of KAW turbulence in three spatial dimensions. A magnetic antenna potential amplitude, evolved in time according to a Langevin equation~\cite{tenBargecpc14}, is externally prescribed at the largest scale of the simulation to model the energy injection at the outer scales of the system.
The driven modes in units of the lowest wave numbers in	$(k_x, k_y, k_z)$
are $(0, 1,\pm 1)$ and $(1,0,\pm 1)$. The modes are driven using a mean antenna frequency $\omega_a = 0.9 \, \omega_{A0}$, decorrelation rate $\gamma_a = 0.7 \, \omega_{A0}$, 
and amplitude $A_{\parallel,0} = \omega_{A0} B_0 / (\sqrt{2} \, k_{\perp,0}^2 v_{A})$, where $\omega_{A0}$ is the frequency of the slowest Alfv\'en wave in the system. Compared to Ref.~\cite{toldprl15} we filter out the antenna wavenumber modes when computing our diagnostics. We chose physical parameters of the simulation to be close to the solar wind conditions at 1 AU, with $\beta_{i}=8\pi n_{i}T_{i}/B_{0}^{2}=1$ and $T_{i}/T_{e}=1$. Proton and electron species are included with their real mass ratio of $m_{i}/m_{e}=1836$. Electron collisionality is chosen to be $\nu_{e}=0.06\, \omega_{A0}$ (with $\nu_{i}=\sqrt{m_{e}/m_{i}}\nu_{e}$). The evolution of the gyrocenter distribution is tracked on a grid with resolution $(N_{x}, N_{y}, N_{z}, N_{v_{\parallel}}, N_{\mu},N_{\sigma})=(768, 768, 96, 48, 15, 2)$. The resolution covers a perpendicular wave number range of $0.2\le k \rho_{i}\leq 51.2$ (or $0.0047\leq k_{\perp} \rho_{e}\leq1.19)$. Here, $\rho_\sigma= \sqrt{T_{\sigma} m_{\sigma}} c / e B_0$ is the species gyroradius and $\sigma$ denotes the species type. The range of wave numbers covered corresponds to a perpendicular box size of $L_{\perp} = 10 \pi \rho_i$ (the equivalent of roughly $44$ ion inertial lengths). In the dynamically fully developed state, the turbulent fields display strong intermittent fluctuations, which can be examined by calculating the probability density function (PDF) of field increments for various lags in directions perpendicular to the mean magnetic field. In Fig.~\hyperref[structure]{\ref*{structure}}, we analyze the PDFs of magnetic field increments $\delta B_y(l) = B_y (\vec{r} + \vec{l}) - B_y(\vec{r})$ for different separation lengths $l$. We observe that the increments at large separation approach a Gaussian, while the increments at small separations develop increasingly strong non-Gaussian tails, indicating intermittency~\cite{frisch_book}. 

{\em Plasma heating and dissipation measures.---} Boltzmann's H-Theorem states that any increase in entropy, necessary for irreversible heating, can only occur due to collisions~\cite{howesapj06}. For this reason, in the context of this paper heating refers to the collisional dissipation of GK turbulence, for which the free energy represents a measure of the intensity of turbulent fluctuations. The collisional dissipation rate for each plasma species is defined in real space ($x,y,z$) at a given time $t$ as
\begin{eqnarray}
Q_{\sigma}(x,y,z,t) = \frac{2\pi B_0}{m_{\sigma}} \int  {\rm d}v_{\parallel} {\rm d}\mu \frac{T_{\sigma}}{F_{0,\sigma}} h_{\sigma} \, C[f_{\sigma}].
\end{eqnarray}
Here, $F_{0,\sigma}$ is a Maxwellian background distribution with background density $n_{\sigma}$ and temperature $T_{\sigma}$, $f_{\sigma}$ is the perturbed GK distribution function, and $h_{\sigma}$ its nonadiabatic part. The nonadiabatic part is given by $h_{\sigma} = f_{\sigma} + (q_{\sigma} \bar{\phi}_{1,\sigma} + \mu \bar{B}_{1,\parallel \sigma}) F_{0,\sigma}/T_{\sigma}$, where 
$\phi$ and $B_{1,\parallel \sigma}$ are, respectively, the self-consistent electrostatic and parallel magnetic field up to first order and the over-bar refers to a gyro-average. 
The collision operator $C[f_{\sigma}]$ is of a linearized Landau-Boltzmann~\cite{landau} type with energy and momentum conserving terms (see Supplemental Material \footnote{See Supplemental Material at  [URL will be inserted by publisher],  which includes Refs.~\cite{helander,hazeltine,hauke13,brizard_pop04,xu_91},  for an explicit form of the collision operator implemented in {\sc Gene}} for an explicit form of the collision operator implemented in {\sc Gene}). Finally, the total plasma collisional dissipation rate is obtained as a sum of contributions from the two species ($Q=\sum_\sigma Q_{\sigma}$). For the parameters used in this work, the electron collisional dissipation peaks in the interval $1<k_{\perp} \rho_i<10$, while the ion collisional dissipation peaks in the $10<k_{\perp} \rho_i<50$ interval. More importantly, about $70 \%$ of the total collisional dissipation is found to arise from electron collisions.

The perturbed part of the distribution function is assumed small here ($f_{\sigma} \ll F_{0,\sigma}$). Hence, collisions are accurately described by a linearized collision operator within this approximation, whereas large fluctuations in $f_{\sigma}$ would require the use of a nonlinear operator~\cite{pezzi_jpp15,pezzi_prl16}. While neglecting large fluctuations, our model still captures an essential feature of (weakly) collisional dissipation, that is, heating is increased locally due to large velocity gradients in the perturbed distribution function. As small velocity scales develop naturally in the weakly collisional limit through kinetic effects, such as linear Landau damping and nonlinear phase mixing, the small scale velocity gradients contribute more and more to the heating of the plasma.
It is also worth emphasizing that the collision frequency for this study was chosen sufficiently low so as not to limit the kinetic range physics contained in the GK model. Therefore, collisionality should be understood as an ultimate sink for free energy that prevents unlimited filamentation of structures in velocity space while still allowing for a broad range of kinetic effects.  

We compare the local heating rates to the so-called electron frame dissipation measure (EFDM)~\cite{Zenitani2011}
\begin{eqnarray}
D_e  = \mathbf{J} \cdot  ( \mathbf{E} + \mathbf{v}_e\times \mathbf{B}),
\end{eqnarray}
where $\mathbf{v}_e$ the electron fluid velocity. The standard definition of $D_e$ contains a second term proportional to the charge density, which 
is cancelled out by the quasineutrality assumption made here. The EFDM is related to the work done by the non-ideal part of the electric field in the generalized Ohm's law ($\propto J^2$ in resistive MHD) and it is the same in the electron and ion frame when quasineutrality is assumed~\cite{wanprl12}. Several authors have recently considered $D_e$ as a proxy for turbulent dissipation in situations where no explicit expression for collisional dissipation is 
available~\cite{sundkvistprl07, wanprl12, wanprl15}. Hereafter, we will consider only $D_e >0$ since this is where the energy is taken out of the electromagnetic fields.
\begin{figure}[!htb]
\centering
\includegraphics[width = 0.47\textwidth]{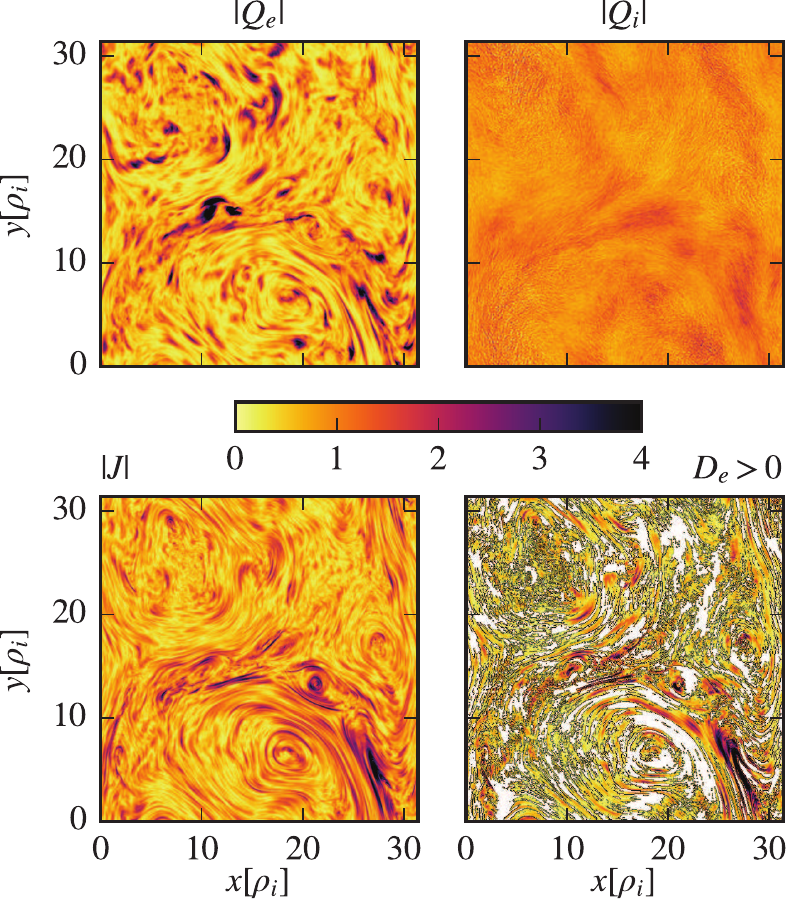} \\
\caption{(Color online) Magnitude of the electron and ion heating density, current density, and the EFDM for a given $z$ plane at a given instant in time. The fields have been normalized to their RMS values and values exceeding the range $[0,4]$ have been clipped to the bounding values of the chosen color (gray) scale.}
\label{fig_QJD_a}
\end{figure}
\begin{figure}[!htb]
\centering
\includegraphics[width = 0.47\textwidth]{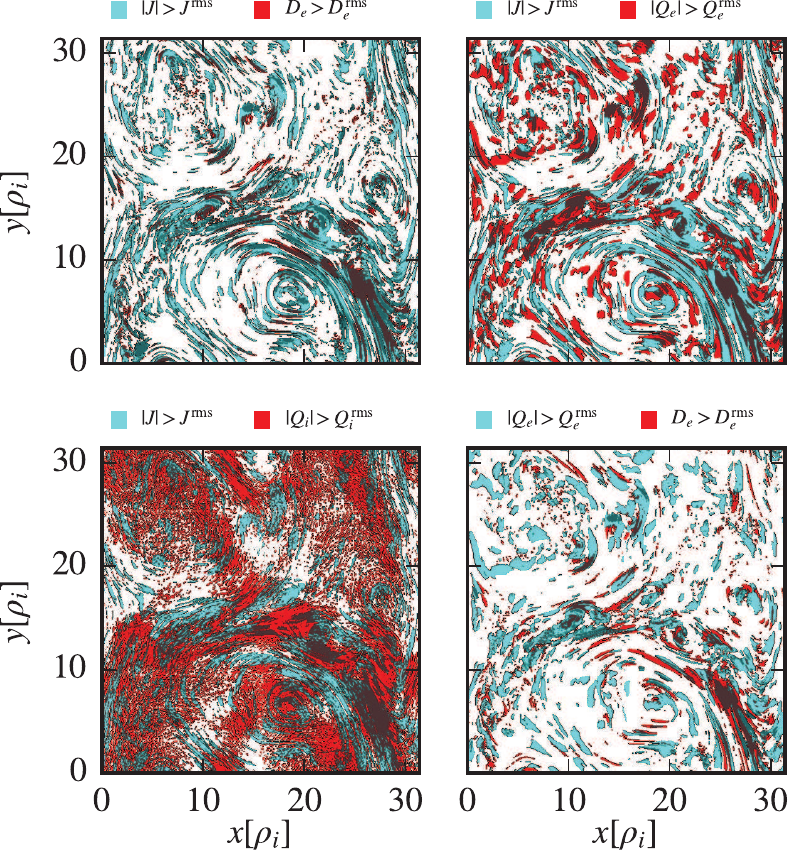} \\
\caption{(Color online) Superposition of two different fields for values larger than their RMS values. Matching 
values are shown with dark color (gray).}
\label{fig_QJD_b}
\end{figure}

{\em Relation of current sheets to heating and the EFDM.---} We observe that the current is distributed in a hierarchy of structures: smooth sheets that are twisted in large scale vortices coexist with smaller scale filaments (Fig.~\hyperref[fig_QJD_a]{\ref*{fig_QJD_a}}). The real space structure of ion and electron heating differs. While $Q_i$ is more uniform, $Q_e$ clearly shows  a patchy nature. For a given $z$  plane at a given instant in time, we define current sheets as regions of high current that exceed the space-averaged root-mean-square (RMS) value $J^{rms}=\langle J^2 \rangle^{1/2}$. Similarly, we look at regions of high  heating rates, for which $|Q_\sigma(x,y)|\ge Q_\sigma^{rms}$. For electrons, these structures contribute to about $50\%$ of the total electron heating. Already this fact alone demonstrates that a significant amount of the electron heating cannot be attributed to regions with intense peaks in the heating, even though a patchy nature for $Q_e$ is evident from Fig.~\hyperref[fig_QJD_a]{\ref*{fig_QJD_a}}. The amount of electron heating contained in current sheets is just $35 \%$ of the total space-integrated value. The same comparison for $Q_i$ yields $30 \%$ and
$65 \%$ for $D_e > 0$.  Restricting the analysis to intense heating structures above the RMS value (Fig.~\hyperref[fig_QJD_b]{\ref*{fig_QJD_b}}), we find that only $45 \%$ of $Q_e$ and $35 \%$ of $Q_i$ peaks are in current sheets.  Not only is the  heating not dominated by intense value structures, but these structures fall mostly outside current sheets. The same analysis for $D_e$ yields $90 \%$. These results agree with previous findings, where it was shown that EFDM was predominantly 
concentrated in current sheets~\cite{wanprl12, wanprl15}. Regarding the comparison of $D_e$ and $Q_e$, only $25 \%$ of high $Q_e$ values match the high values of $D_e$.  We have also verified that the agreement between $Q_e$ and $D_e$ does not significantly  improve by choosing a lower threshold to define intense peaks in $D_e$. 

\begin{figure}[!htb]
\centering
\includegraphics[width = 0.47\textwidth]{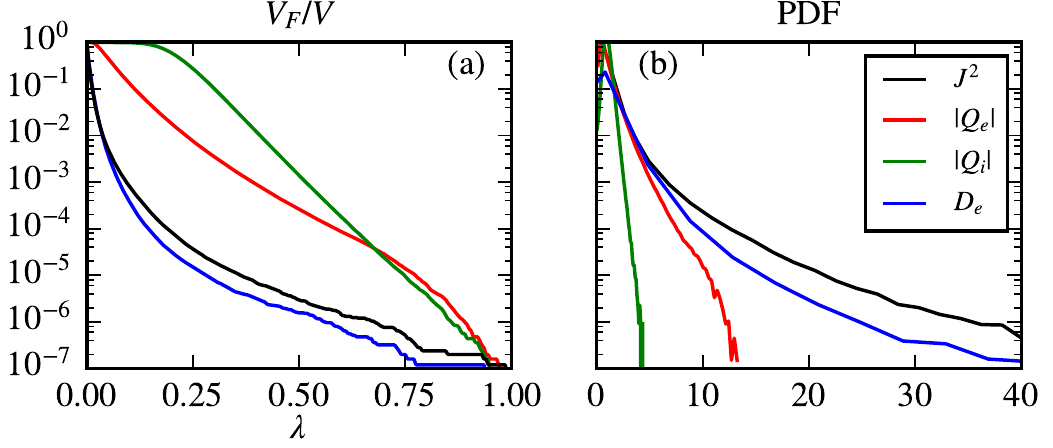}
\caption{(Color online) (a) Volume filling ratio (see text) and (b) PDFs comparing the behavior 
of $|Q_e|$, $|Q_i|$, and $D_e$. The $J^2$ curves are displayed for reference, showing a good agreement with $D_e$.}
\label{conditional}
\end{figure}

Finally, to better gauge how the EFDM and heating structures are distributed in space and in current sheets we employ a series of  diagnostics, similar to the ones introduced in Ref.~\cite{wanprl12} and employed for the analysis of $D_e$.  For a field $F$ with a total volume $V$, we define the volume $V_{F}$ as the volume occupied by values of $|F|$ larger than $\lambda F^{\max}$, where $F^{\max}$ is the maximal absolute value of $F$ at any $(x,y,z)$ point and $\lambda\in[0,1]$. In Fig.~\hyperref[conditional]{\ref*{conditional}(a)} we plot the volume filling ratios $V_{F}/V$ versus $\lambda$. We see that high intensity heating structures (with respect to their maximal value) occupy a much larger volume than $D_e$, the latter showing a  very good agreement with the filling ratio of  $J^2$. Fig.~\hyperref[conditional]{\ref*{conditional}(b)} compares the PDFs of $|Q_e|$, $|Q_i|$, $D_e$, and $J^2$. $D_e$ has the broadest distribution, which is again closer to the distribution of $J^2$, while $Q_e$ and $Q_i$ have a much narrower distribution. These results clearly show that the EFDM cannot be considered as representative for turbulent heating in weakly collisional plasmas. An explanation  can be given in terms of the cascade through phase space. $D_e$ represents only an injection of electromagnetic energy into the cascade and agrees very closely with the locations of current sheets. The injection of energy and (irreversible) dissipation generally form a causal relationship. Energy transfer from fields to particles is necessary for dissipation to occur, however, when collisions are rare, the conversion of free energy into heat need not happen at the same time as the actual transfer. In particular, the fluctuations are advected through phase space from the point where energy transfer occurs to the point where velocity gradients are sufficiently large for the collisional term to become significant. Once the free energy reaches the smallest scales at a later time, it is no longer confined to current sheets. For solar wind turbulence, our result could be modified when considering the role of inertial  range dynamics. In a simulation with a better resolved inertial range, the increased level of intermittency may give rise to more pronounced non-Maxwellian features around the most intense current sheets. These features are beyond the scope of the present work, where fluctuations in the  distribution function are assumed to be small  and the simulation box covers only the tail of the inertial range.

\begin{figure}[tb]
\centering
\includegraphics[width = 0.4\textwidth]{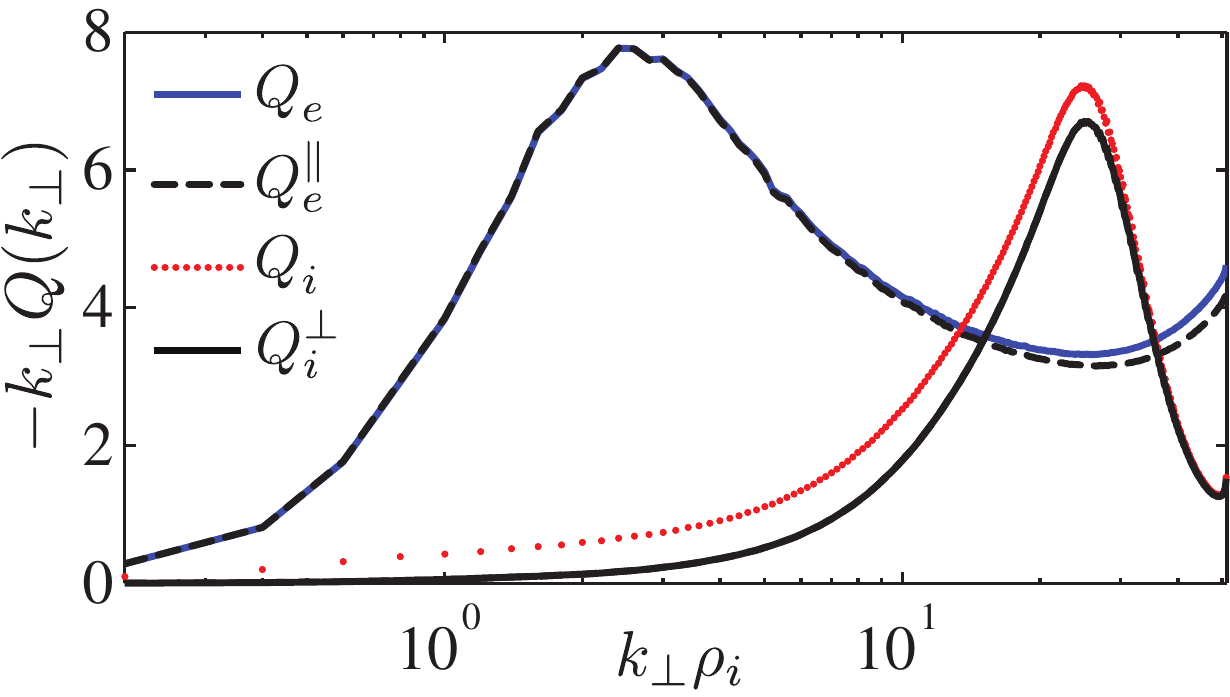}
\caption{(Color online) 
Normalized perpendicular wavenumber spectra for electron and ions heating rates. The parallel electron and  perpendicular ion contributions to the heating rate are shown for comparison. Curves are multiplied by $k_{\perp}$ so the area under the curve (in log scale for $k_{\perp}$) is proportional 
to the collisional dissipation rate.}
\label{fig03}
\end{figure}

{\em Plasma heating channels.---} As shown above, the real space structure of ion and electron heating differs. To gain insight into the heating route preferred for each species, we split the collisional dissipation into the following contributions: 
\begin{eqnarray}
Q_{\sigma} = Q^{\parallel}_{\sigma} + Q^{\perp}_{\sigma} + Q^{\rm coupling}_{\sigma},
\end{eqnarray}
where $Q^{\parallel}_{\sigma}$ involves only the parallel velocity derivatives of $C[f_{\sigma}]$, $Q^{\perp}_{\sigma}$ involves only the magnetic moment $\mu$ derivatives and contains a $[k_{\perp} \rho_\sigma]^2$ spatial term due to the GK transformation of the particle perpendicular velocity, and $Q^{\rm coupling}_{\sigma}$ is a coupling term which involves the product of both parallel and perpendicular velocity derivatives (see Supplemental Material~\cite{Note1}). As the collisional term $C[f_{\sigma}]$ has the same form for both ions and electrons, observing a dominance of $Q^{\parallel}_{\sigma}$ or $Q^{\perp}_{\sigma}$ implies  that smaller parallel or perpendicular velocity structures are preferentially developed. The development of parallel velocity structures is due to linear phase mixing and indicates a predilection for Landau damping. The results are shown in Fig.~\ref{fig03}. The electron collisional dissipation ($Q_e$) spectrum peaks at low $k_{\perp}$. This hints that Landau damping is the preferred route for electron heating~\cite{TenBargeapj13}, for which linear phase mixing is crucial, as discussed in Refs.~\cite{Loureiro2013, Numata2015} for weakly collisional reconnection in two dimensions. For Alfv\'enic turbulence addressed in the present Letter, the importance of electron Landau resonance and linear phase mixing is clearly shown by the fact that $Q_e$ is dominated by the $Q_e^{\parallel}$ contribution (approximately  $96\%$ of the total electron heating), a result possible only due to the presence of strong parallel velocity gradients. By comparison, the ion free energy is cascaded to small perpendicular scales and dissipated around $k_{\perp} \rho_i\! \sim \!25$. Approximately $80 \%$ of the total ion free energy is dissipated solely due to the $Q_i^{\perp}$ contribution. The main ion heating channel is therefore nonlinear phase mixing, the latter being characterized by the simultaneous generation of small perpendicular velocity and spatial structures~\cite{schekochihinajss09}, in contrast to linear Landau damping which effects the parallel velocity structure. 

{\em Discussion and conclusions---}
In this Letter, we analyzed the relation between turbulent heating rates and current sheets in a weakly collisional plasma by comparing directly 
both quantities in real space. Using gyrokinetic simulations of KAW turbulence, we obtained  evidence that the locations of current sheets (defined as regions of high current) do not 
generally correspond to peaks in the electron and ion heating rates. Several authors have previously argued that heating in the solar wind occurs mainly in current sheets. However, one needs to take into account that these analyses were performed  either by using various dissipation  proxies instead of the actual local heating rates~\cite{parasharpop11, TenBargeapj13, wanprl12, Karimabadipop13, wanprl15} or without measuring directly the heating rate in real space~\cite{TenBargeapj13}.  Indeed, when comparing $D_e$ with the electric current, we find a good agreement between these quantities, but not between $J$ (or $D_e$) and the electron and ion heating rates. Furthermore,  we showed that plasma heating is highly anisotropic in velocity space, providing first time direct measurements that identify the dominant collisional direction for each species. We identified Landau damping as the preferred route for the electron heating, while nonlinear (perpendicular) phase mixing is the channel responsible for ion heating. These results demonstrate the importance of kinetic dynamics and the use of well-defined collisional dissipation measures for studying plasma heating, further stressing the need for in-situ spacecraft measurements that would allow precise estimates of 
collisional heating from particle distribution functions.

\begin{acknowledgments} 
{\em Acknowledgements.---} We would like to thank Gabriel Plunk and Silvio Cerri for fruitful discussions. We also acknowledge the Max-Planck Princeton Center for Plasma Physics for facilitating the discussions that lead to this paper. The research leading to these results has  received funding from the European Research Council under the European Unions Seventh Framework Programme (FP7/2007V2013)/ERC Grant Agreement No. 277870. The gyrokinetic simulations presented in this work used resources of the National Energy Research Scientific Computing Center, a DOE Office of Science User Facility supported by the Office of Science of the U.S. Department of Energy under Contract No. DE-AC02-05CH11231.
\end{acknowledgments}

%

\end{document}